\documentclass[sigconf]{acmart}

\usepackage{color}
\usepackage{ifthen}
\usepackage{booktabs}
\newboolean{showcomments}
\setboolean{showcomments}{true} 
\ifthenelse{\boolean{showcomments}}
    {
        \newcommand{\per}[1]{\textcolor{blue}{{\it [Per says: #1]}}}
        \newcommand{\johan}[1]{\textcolor{cyan}{{\it [Johan says: #1]}}}
    }
    {
        \newcommand{\per}[1]{}
        \newcommand{\johan}[1]{}
    }

\AtBeginDocument{%
  \providecommand\BibTeX{{%
\normalfont B\kern-0.5em{\scshape i\kern-0.25em b}\kern-0.8em\TeX}}}

\begin{document}

\title{Public Sector Platforms going Open: Creating and Growing an Ecosystem with Open Collaborative Development}


\author{Johan Lin{\aa}ker}
\email{johan.linaker@cs.lth.se}
\author{Per Runeson}
\email{per.runeson@cs.lth.se}
\affiliation{%
  \institution{Department of Computer Science, Lund University}
  \streetaddress{Ole R\"{o}mers väg 3}
  \city{Lund}
  \country{Sweden}}

\renewcommand{\shortauthors}{Lin{\aa}ker and Runeson.} 


\begin{abstract}
\textit{Background}: By creating ecosystems around platforms of Open Source Software (OSS) and Open Data (OD), and adopting open collaborative development practices, platform providers may exploit open innovation benefits. However, adopting such practices in a traditionally closed organization is a maturity process that we hypothesize cannot be undergone without friction. 

\textit{Objective}: This study aims to investigate what challenges may occur for a newly-turned platform provider in the public sector, aiming to adopt open collaborative practices to create an ecosystem around the development of the underpinning platform.

\textit{Method}: An exploratory case-study is conducted at a Swedish public sector platform provider, which is creating an ecosystem around OSS and OD, related to the labor market. Data is collected through interviews, document studies, and prolonged engagement.

\textit{Results}: Findings highlight a fear among developers of being publicly questioned for their work, as they represent a government agency undergoing constant scrutiny. Issue trackers, roadmaps, and development processes are generally closed, while multiple channels are used for communication, causing internal and external confusion. Some developers are reluctant to communicate externally as they believe it interferes with their work. Lack of health metrics limits possibilities to follow ecosystem growth and for actors to make investment decisions. Further, an autonomous team structure is reported to complicate internal communication and enforcement of the common vision, as well as collaboration. A set of interventions for addressing the challenges are proposed, based on related work.

\textit{Conclusions}: We conclude that several cultural, organizational, and process-related challenges may reside, and by understanding these early on, platform providers can be preemptive in their work of building healthy ecosystems.
\end{abstract}





\maketitle

\section{Introduction} 
The creation of ecosystems around software and data platforms is a commonly adopted approach to exploit the benefits of open innovation~\cite{zuiderwijk2014innovation, munir2015open}. By enabling the development of third-party applications on top of the platform, new niches can be addressed, the market expanded and so also the adoption of the platform~\cite{jansen2013defining}, and new technologies~\cite{jansen2019managing}. By also releasing the platform under an open license and adopting an open collaborative approach for the platform development, the platform provider can benefit from external contributions and thereby potentially accelerate both innovation and development while sharing the costs and risks with the ecosystem~\cite{munir2015open, runeson2019open}. 

The platform may consist of Open Source Software (OSS)~\cite{alves2017software}, Open Data~\cite{zuiderwijk2014innovation}, or both~\cite{runeson2019open}. Independently of which, creating an open healthy ecosystem~\cite{kilamo2012proprietary, jansen2014measuring} and switching to an open and collaborative development model is for newly-turned platform providers a maturity process in terms of culture, processes and organizational structures~\cite{jansen2020focus}. We hypothesize that this transformation is not without friction and challenges along the way. 

The research goal of this study is to shed light on: \begin{quote}\textit{what challenges a platform provider may face when adopting open collaborative practices to create and grow an ecosystem around the underpinning platform}.\end{quote} This investigation is done through a case study~\cite{runeson2012casestudy} of the Swedish Public Employment Service (SPES), a public agency responsible for enabling match-making between job-seekers and employers on the Swedish labor market. SPES is in the early phases of growing an ecosystem (JobTech Dev\footnote{https://jobtechdev.se/en}) of actors, with the vision to improve the digital match-making and guidance services by providing a platform of Open Government Data (OGD), related APIs and OSS. As part of this effort, they want to collaborate with the ecosystem on the development and enrichment of both the OGD and OSS.

Seven semi-structured interviews are conducted with a questionnaire, based on Jansen's maturity model for software development governance in the context of software ecosystems~\cite{jansen2020focus}. Interview data is supplemented with documentation analysis and field notes from the first author who is embedded as an action researcher in the unit of SPES, which is responsible for the platform development. 

The case study is a part of the first and diagnosing phase of an action research project~\cite{staron2019action} focused on SPES and its aim of growing a successful and healthy ecosystem of OGD and OSS. The identified challenges provide a foundation for action planning, i.e., the design of interventions to be introduced, which will be part of the future work.

The rest of this paper is structured as follows. In section~2, a background is presented on ecosystems and open collaborative development practices, followed by the study's research design in section~3. Results from the study are presented in section~4, including an overview of current practices, adopted by SPES, as well as identified challenges. Results are discussed in section~5 where we also propose a number of interventions to address the identified challenges. Finally, we conclude the case study in section~6.

\section{Background}\label{sec:rw}
In this section we present background research on ecosystems and practices that can be adopted to create and grow a healthy ecosystem around software and data. We further provide an overview of the open and collaborative development practices used in OSS ecosystems.

\subsection{Creating and Growing an Ecosystem}\label{sec:rw:ecosystem}
The ecosystem metaphor originates from the biological domain and has been applied both to business in general~\cite{iansiti2004keystone} and to the more specific domains of software~\cite{jansen2013defining, manikas2016revisiting} and data~\cite{zuiderwijk2014innovation, oliveira2019investigations}. In this study, we consider the case where the platform consists of both software and data, and more specifically OSS and OGD. We consider an ecosystem as: \begin{quote} \textit{a networked community of organizations, which base their relations to each other on a common interest in an underpinning technological platform consisting of OGD, OSS, and open standards, and collaborate through the exchange of information, resources and artifacts};\end{quote} a definition adapted from Zuiderwijk et al.~\cite{zuiderwijk2014innovation} and Jansen and Cusomano~\cite{jansen2013defining}.

The platform provider is usually the one orchestrating the ecosystem and setting the governance and governance structure, i.e., deciding on processes and procedures for the use and development of the platform, as well as the distribution of rights and responsibilities among the members of the ecosystem~\cite{baars2012framework}. Designing the right governance is therefore pivotal to creating and growing a healthy ecosystem~\cite{jansen2013defining}. A healthy ecosystem is one that shows an ability to \textit{``endure and remain variable and productive over time''}~\cite{manikas2013reviewing}. By measuring the health, the platform provider can take necessary actions in terms of adapting the ecosystem governance and platform development to preserve or improve the health of the ecosystem~\cite{jansen2014measuring}. This can be achieved, for example, by \textit{``creating/refining rules and processes for the actors, communicating plans to the actors (for example, by road-mapping), organizing the ecosystem development through, for example, release management, making changes to the platform and other software components, changing the revenue model for internal products, and controlling the actor population and motivation by modifying the model by which the actors participate in the ecosystem''}~\cite{manikas2013software}.

Jansen~\cite{jansen2020focus} proposes a maturity model to help platform providers assess and advance their software ecosystem governance practices in order the grow a healthy ecosystem~\cite{jansen2014measuring}. The maturity model covers a wide range of aspects but in this study, we are mainly interested in the focus area of \textit{Software Development Governance} and the connecting eight practices (denoted P\# below) associated with co-development of the platform with the ecosystem. The focus area highlights practices enabling ecosystem members to improve the \textit{app testing (P1)} and as well as the \textit{application quality (P2)}, for example by sharing of test data and quality issues. 
\textit{Developer relationships (P3)} should at a minimum be maintained with informal contacts and meetups, while more mature practices focus on enabling the ecosystem's members to help each other and collaborate directly.

\textit{Developer support (P4) and process automation (P5)} focus on enabling ecosystem members to get started with their application development. At a minimum, informal support should be provided along with necessary documentation and optimized installation procedures for any supporting software. More mature practices highlight the use of a more collaborative approach, for example, with a common ticketing system, and open process for road-mapping and development of the platform and any extensions.

\textit{Sharing the requirements (P6) and roadmap (P7)} of the platform should at least be done on an informal level. More mature practices highlight a more open approach through the use of requirements portals and open roadmaps, but also involving the ecosystem in the requirements process where they can actively assert, analyze and prioritize the requirements, both short and long-term.

The eight and last practice listed~\cite{jansen2020focus}, \textit{developer monitoring (P8)} is focused on practices for collecting needs and wishes from the ecosystem regarding platform functionality, as well as to ecosystem governance and development processes and infrastructure. The monitoring should at least be done at an informal level, while more mature levels require documentation and automated data collection.

The more mature practices have in common that they highlight openness and transparency in the development process, taking inspiration from how co-development is performed in OSS ecosystems~\cite{alves2017software, jansen2013defining}. Platform providers can, therefore, take inspiration from OSS development practices and consider how OSS ecosystems specifically are created and maintained~\cite{kilamo2012proprietary}. As highlighted by Runeson~\cite{runeson2019open}, similar practices may potentially also be applicable in the co-development and enrichment of open data in open data ecosystems~\cite{oliveira2019investigations}.


\subsection{Development and Governance in Open Source Software Ecosystems}
Although there is no single model for the collaborative development practices used in OSS ecosystems~\cite{feller2002understanding} (or communities as they are commonly referred to~\cite{munir2015open}), these practices may often be described as informal compared to \textit{``the usual industrial style of development''}~\cite{mockus2002two}. The coordination processes may range from ad-hoc to more structured, based on the size and complexity of the OSS project, as well as the traditions of its ecosystem~\cite{ernst2012case, mockus2002two}. The processes and supporting infrastructure need to facilitate a distributed and decentralized ecosystem of individuals and organizations, working asynchronous and in parallel, often on self-assigned tasks based on interest and agenda~\cite{mockus2002two, feller2002understanding}. 

Recent research on barriers for newcomers and episodic contributors to OSS development also highlight the social and cultural aspects~\cite{steinmacher2019overcoming, barcomb2020managing}, for example, the need to respond quickly and friendly to queries, make newcomers feel welcome and appreciated, be considerate of potential cultural differences as well as the level of knowledge, and also provide mentorship and resources for onboarding and high-quality documentation.

Considering the software development process, requirements are commonly asserted (rather than elicited)~\cite{castro2012differences}, analyzed and evolved in overlapping fragments across the different tools used for communication and collaboration within the ecosystem~\cite{ernst2012case, german2003gnome}, for example, as issues in an issue tracker, threads in mailing-lists, or commits in the version control system~\cite{scacchi2002understanding}. Bugs are commonly reported, discussed and managed similarly alongside, and sometimes also intertwined with the requirements~\cite{laurent2009lessons}. Contributions are peer-reviewed to maintain quality assurance and coding standards~\cite{feller2002understanding}. All discussions and contributions are persisted transparently for anyone to review, for example, decision rationale or software functionality~\cite{alspaugh2013ongoing,german2003gnome}.

Decisions on what to accept and the general direction of the OSS project is usually determined by a core group of individuals overseeing the project management~\cite{nakakoji2002evolution}, although still considering the will of the ecosystem~\cite{laurent2009lessons, aagerfalk2008outsourcing}. The level of influence on these types of decisions is commonly gained by building trust and proving merit through active contributions (technical and non-technical) and symbiotic relationships~\cite{linaaker2019community,dahlander2005relationships}. What makes up a merit is highly context-dependent~\cite{eckhardt2014merits, o2007emergence} and how it affects one's progression in influence and responsibility is dependent on the type of governance~\cite{de2007governance}. In OSS ecosystems with autocratic governance-style, power is centered around a single organization or individual, while in OSS ecosystems with democratic tendencies the power is decentralized and distributed among the members of the ecosystem~\cite{de2013evolution}.

Companies' involvement in an OSS ecosystem is dependent on its importance for business~\cite{dahlander2005relationships, linaaker2019community}, along with contextual factors such as the ecosystem's maturity and need for contributions, and in what ways the organization can contribute~\cite{butler2018investigation, linaaker2019community}. Contributions can be made by a company's employees or subcontracted individuals, usually with a central position inside an OSS ecosystem~\cite{butler2019company}. Financial support, knowledge-sharing, and board-memberships are some other ways in how companies can contribute in order to reach their goals while also supporting the ecosystem~\cite{linaaker2019community}.

OSS development practices can also be applied internally within an organization and are then referred to as inner source~\cite{stol2014key}. Universal access to code and documentation, common infrastructure, and peer-review are some of the practices commonly applied~\cite{linaaker2014on}.


\section{Research Design}
We launched an exploratory case study~\cite{runeson2012casestudy} to investigate the practices used by a platform provider in terms of collaborating with its ecosystem, and what challenges they face in this process. Below we first describe the studied case, followed by an overview of the research methods used in this study. 

\subsection{Case Presentation} 
The case is a platform and development project initiated by the Swedish Public Employment Service (SPES), a public agency responsible for enabling match-making between job-seekers and employers on the Swedish labor market. SPES has an IT division of about 600 employees, but this study focuses specifically on a unit responsible for the growing and facilitation of the JobTech Dev ecosystem and its underpinning platform of OGD and OSS. This unit, or JT-unit as we refer to it below, has the ambition to adopt an open development model for its platform in order to involve and collaborate with the ecosystem members. They wish to take inspiration from how OSS communities work and hope that members will make contributions related to both data and software. The unit of analysis of this study, therefore, is constituted by \emph{the practices used by the JT-unit for collaborating and interacting with the ecosystem on the development of the underpinning platform}.

The platform consists of four main parts, also referred to as products:

\begin{itemize}
    \item \textit{Jobs} is a data-set of job-advertisements on the Swedish labor market, offered through a central API. The ads are collected from several ad-platform-providers.
    \item \textit{Career} is a service that allows citizens to store résumé data and provide consent to share with, for example, recruitment and staffing firms, and government agencies. The consent to share the data with any organization can be revoked at any time.
    \item \textit{Taxonomy} is a data-set of skills, job titles, and relationships in between that enable actors related to the Swedish labor market to speak the same language concerning, for example, job-ads, résumés, statistics, and reports.
    \item \textit{Search} is a service that allows users to search among the job-ads currently provided by SPES. The search engine is available both as an OSS project (based on Elasticsearch) and through APIs.
\end{itemize}

The ecosystem surrounding the platform consists of actors, such as staffing and recruitment companies, education providers, public entities, trade unions, employer associations, and insurance providers. The size varies between small startup companies to large national and international organizations. Their interest in the platform may be limited to certain parts depending on context and business. The actors can, therefore, be viewed as stakeholders either to the platform in general or to a specific subset of it.

The JT-unit has about 30 employees divided into seven different teams. They use a flat team structure, meaning that everyone within the unit's organization belongs to a team and is responsible for a certain product or function related to the platform, for example, development operations, website, and communication, and team support. Most teams are centered in Stockholm, while some are divided and dispersed geographically within Sweden.

The overall strategy and planning for the platform are overseen by a product manager. However, each team is autonomous and responsible for its respective products or function. This responsibility includes development, related infrastructure and software engineering processes, and how to open up for collaboration with the ecosystem's members.

\subsection{Research Methods}
\label{researchMethods}
\begin{figure}[tbp]
\centering
\includegraphics[width=0.8\columnwidth]{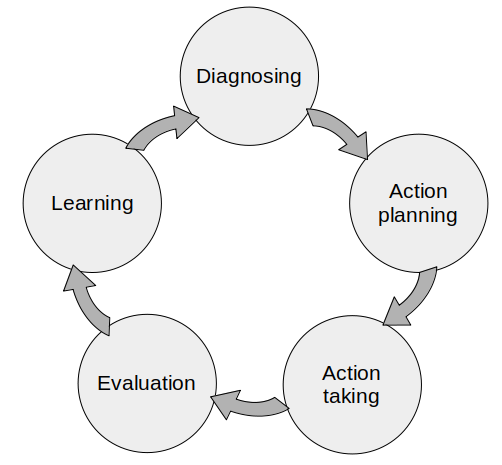}
\caption{The five steps of the Canonical Action Research cycle (adopted from Staron~\cite{staron2019action}) starting with the Diagnosing phase which is the context for this study.}
\label{fig:actionresearchcycle}
\end{figure}

The study is part of the diagnosing phase of an action research project~\cite{staron2019action} aimed at supporting the unit in growing a healthy ecosystem, see Fig. \ref{fig:actionresearchcycle}. The identified challenges and suggested interventions will provide input to the second and following phase of action planning. 

The first author has therefore been embedded as an action researcher in the unit for six months before the reporting of this study. During this period, the first author has collected documentation and made extensive field notes describing the unit's ongoing work and challenges. The second author primarily acts as an academic advisor in the overarching action research project.

The first researcher has mostly interacted with the unit remotely with sporadic visits, in line with the generally decentralized working approach of the unit. Peer-debriefings with the second author has been performed to minimize researcher bias~\cite{easterbrook2008selecting}.

\begin{table}[t]
\caption{Roles of interviewees and their specific area of focus.}
\label{tbl:interviewees}
\begin{tabular}{@{}ll@{}}
\toprule
\textbf{Interviewee Role} & \textbf{Area of focus} \\ \midrule
Product Manager & General \\
Community Manager & General + Career \\
Business Analyst & General + Jobs \\
Developer & DevOps and Infrastructure \\
Developer & Search and Jobs \\
Developer & Career \\
Developer & Taxonomy \\ \bottomrule
\end{tabular}
\end{table}

Seven semi-structured interviews were conducted to supplement and triangulate with the informally collected data, to improve the internal validity~\cite{runeson2012casestudy} of the study. To ensure construct validity~\cite{runeson2012casestudy}, the interview questionnaire (see Appendix~\ref{app:questionnnaire}) was based on a focus area maturity model for platform providers that can help to assess and improve their practices for growing a healthy software ecosystem. The focus area of software development governance, as presented in section~\ref{sec:rw:ecosystem}, was used specifically in the construction of the questionnaire.

The set of interviewees (see table~\ref{tbl:interviewees}) include the product manager, community manager, and business analyst, whose responsibilities include the whole platform and ecosystem development. Additionally, four developers were interviewed from the platform's main products, as well as the team responsible for the development and operational support to the other teams. The interviewees are not identified by any markers nor attributed in the reporting of the results due to integrity and ethical reasons, as this would enable the tracing between statements and interviewees, primarily by others within the organizations of the JT-unit and SPES at large.

Each interview was performed by the first author and audio-recorded with notes taken. Notes were inductively coded into two parts, current status and challenges. The first part describes the current development process and infrastructure used, and to what extent collaboration with the ecosystem is performed (see section~\ref{practices}). Codes were then synthesized and grouped under four themes:

\begin{itemize}
    \item Availability of Code and Project Artifacts
    \item Interaction with the Ecosystem
    \item Facilitation and Guidance of the Ecosystem
    \item Requirements Engineering
\end{itemize}

The second part of the codes highlights any challenges perceived by the interviewees, rendering in eight challenges (see section~\ref{challenges}). The challenges were presented and discussed during a workshop with the unit to allow for collective member-checking~\cite{easterbrook2008selecting} and general feedback.

Based on the identified challenge and literature as presented in section~\ref{sec:rw}, we propose a number of interventions linked to the challenges (see section~\ref{discussion}). These interventions together with problem understanding illustrated by the identified challenges constitute the outcome of the first part of the action research cycle as presented in Fig.~\ref{fig:actionresearchcycle}. These results will provide input to the next phase of action planning~\cite{staron2019action}, i.e., the design of interventions to be introduced as parts of future work. We, therefore, wish to highlight that the data was gathered before any intervention has been introduced.

\section{Results}\
Below we present practices used in section~\ref{practices}, both on a team and unit-level, specifically in regards to what extent the ecosystem's members can engage themselves. This is followed by an overview of challenges in section~\ref{challenges}, identified in terms of the practices, currently employed, that may need to be adapted to foster the ecosystem to collaborate on the platform development. The results are based on the inductive coding process of the gathered data as described in section~\ref{researchMethods}.

\subsection{Collaborative Infrastructure and Practices}
\label{practices}


We divide and present the practices and infrastructure used by the JT-unit into four areas: the availability of code and project artifacts; interaction with the ecosystem; facilitation and guidance of the ecosystem; and finally the requirements engineering for the platform.

\vspace{3mm}
\noindent\textit{Availability of Code and Project Artifacts. }
Each team is required to publish any source code related to the platform's APIs as OSS. The choice of tool for sharing the source code is, however, free for teams to decide and differs between GitLab and GitHub. The use of these tools is generally limited to the source code repository functionality with related documentation. The level of detail of the documentation however varies. Some teams are focused on optimizing the time it takes to set up running examples, while others limit themselves to code-level comments and keep the main parts of technical documentation closed. Also, road maps are generally kept closed, as the issue-trackers are actively used in the teams. 

To help ecosystem members getting started, for example applications are provided for the different products. This is especially appreciated by smaller organizations with limited resources.

\vspace{3mm}
\noindent\textit{Interaction with the Ecosystem. }
Ecosystem members can communicate with each other and the teams through Slack -- a synchronous communication channel. The communication channel is primarily used to ask questions which are answered by the concerned team. The JT-unit mainly communicate with each other through an internal instance of Slack.

Members of the ecosystem can also communicate with the JT-unit through a Service Desk, a central first-line support that SPES has set up for all of its services. The teams of the JT-unit make up third-line support in this regard. Tickets are managed in a central and closed ticket-management system. 

Besides the ticket-management system of the Service Desk and Slack, questions and support requests are also submitted via email to individuals within the teams. These may originate from contacts that are established via, for example, meetups and personal contacts. 

Bug reports and feature requests are mixed with technical and support-related questions via all of the aforementioned channels, as well as through the issue-trackers provided by default in GitHub and GitLab, even though these may not be monitored or actively used by each team.

\vspace{3mm}
\noindent\textit{Facilitation and Guidance of the Ecosystem.}
No active facilitation or guidance of the ecosystem is performed via the communication channels. These are primarily used reactively by responding to incoming requests from the ecosystem. Individual meetings with actors for building and maintaining personal contacts and relationships has instead been the main approach for facilitating as well as growing the ecosystem. It is primarily performed by the unit's community manager and product owner who meets up with organizations interested in the platform, as well as existing members of the ecosystem.

Initially, contacts were based on an externally procured stakeholder analysis of all actors related to the Swedish labor market and that might be interested in the ecosystem. The log file of who has fetched an API-key is also used. When the ecosystem started to become more established, most meetings were booked externally by the actors themselves.

Startups are given extra guidance dependent on resource availability as to what APIs are available and to explore certain use cases. The community manager further tries to connect actors where there might be an opportunity for collaborations.

\vspace{3mm}
\noindent\textit{Requirements Engineering. }
The inflow of requirements requested or asserted by the ecosystem is limited, so is any form of active elicitation. An exception concerns key stakeholders, specifically those with large user bases. Dialogues are maintained between concerned teams and these stakeholders to actively elicit requirements. 1-2 day hackathons is one approach used, where developers from concerned team and stakeholder interact and to explore potential use cases.

\subsection{Perceived Challenges with Engaging the Ecosystem}
\label{challenges}

Below we present the seven challenges (labeled CH\#) which we identified through the analysis of the collected data. CH1-2 connect to the possibilities of getting a general overview of the ecosystem population and its health. CH3-4 connect to how the JT-unit facilitates and operates the management of issues and the general requirements engineering process of the platform. CH5-6 relate to the JT-unit's external communication with the ecosystem, while CH7 relates to the internal communication of the unit.

\vspace{3mm}
\noindent\textit{CH1: Metrics of Ecosystem Health.}
From the ecosystem's ''outside'' perspective, it is difficult to get an overview and assess the quality of the software and health of the ecosystem. Information regarding, for example, the number of bug reports or development direction of the software, is closed or spread out. Quality metrics regarding, for example, availability and performance of APIs, are not publicly available. In one case, a contract was signed to ensure an actor in the ecosystem about the JT-unit's commitment to keeping the concerned APIs and data available. The contract may to some degree be compared to a service level agreement (SLA) but more to that of a letter of intent regarding the platform. A perceived risk with the lack of publicly available metrics is that actors may be reluctant to use and base their solutions on the data and software provided by the platform.

\vspace{3mm}
\noindent\textit{CH2: Overview of Ecosystem Population.}
There is currently no overview of actors within the ecosystem and who is using the different parts of the platform. Regarding key stakeholders, this is a tacit knowledge maintained by different individuals in the JT-unit. It is possible to view who has fetched an API-key, but what organizations these belong to and what their use cases are, is not monitored. One perceived risk is that information may be lost when individuals leave the organization. Other perceived risks include important relationships not being maintained, sources of requirements being missed, and misinterpreted views of the ecosystem's health. Compiling and show-casing available use-cases may help to inspire and show the potential of the platform for new and existing members of the ecosystem.



\vspace{3mm}
\noindent\textit{CH3: Reporting and management of issues.}
Since issues, like bugs and feature requests, can be reported through several communication channels, these can risk being missed by the teams. One interviewee reported how (s)he occasionally could identify feature requests when triaging the Service Desk's ticket-management system. Clear instructions for when to use a channel, for what, and how is missing. Even though this may confuse the reporting actor, interviewees also see each channel fulfilling a different purpose. The Service Desk's ticket-management system concerns issues of critical nature and is available 24/7. Questions and issues reported via Slack and mail can not expect to be answered or addressed immediately. This distinction is however not clarified for the ecosystem members. 

Confusion can, however, be noticed internally due to the lack of overview and coordination of questions, support tasks, and feature requests between the different channels. A perceived risk is that these may go unaddressed, to the frustration of the reporting actor.

One interviewee suggests that issues should be categorized based on whether they are relevant for the ecosystem, or only of concern for the teams, and then published on a public and closed issue-tracker, accordingly. The same interviewee mentions that \textit{``double book-keeping may be needed''}, meaning that some issues may have to be listed both on a public issue tracker and a team's internal backlog. The interviewee further highlights that teams work closely with stakeholders within SPES, why internal backlogs may still be needed.

\vspace{3mm}
\noindent\textit{CH4: Elicitation and Prioritization of Requirements.}
Generally, the communication is described as one-way from the JT-unit's perspective towards to ecosystem. This can be observed by the fact that there is no public issue tracker instance and all road-maps containing the long-term planning of the different products are closed in internal infrastructure. Ecosystem members are hence prevented from gaining insights on the direction of the platform without explicitly asking.

The unit has informed the ecosystem about its progress rather than collected requirements from them. One interviewee views this as something positive up until now, as it has allowed the unit to move forward faster. If a dialogue would have been facilitated with the many actors in the ecosystem the interviewee believed that this would have consumed a lot of time. Now, however, a change is noticed where teams are starting to listen more to the ecosystem on what functionality or data-sets are requested.

As an example, the team responsible for Search is reported working towards a more open requirements elicitation and prioritization process. Initially, they have had a close collaboration with a startup which can be considered as an early adopter. They ask users over Slack and at meetups focused on the Search-functionality. Also, they get some input via Service Desk. Meetups are considered valuable as a potential source for gathering requirements from users. 




\vspace{3mm}
\noindent\textit{CH5: External Communication.}
There is a skepticism internally among some of the teams at the JT-unit regarding external communication with ecosystem members and answering questions on communication channels. Some have questioned whether this is a part of their job assignment and whether they should dedicate time to it. A concern is that they may have to spend a day answering questions from the ecosystem. Others in the JT-unit question whether this is a concern as it, in reality, is not that many questions posted and that they believe it to be a natural part of their work. A reason highlighted by one interviewee was that some developers were not that extrovert, neither virtually or physically, which may also be a reason for why the number of meetups has been limited thus far.

Others within the unit, on the contrary, request more meetups and better communication with the ecosystem to showcase the functionality and get feedback or collect feature requests. Some highlight frustration over Slack as a communication channel within the ecosystem. They request a solution where information is openly persisted, structured, and searchable,

Even though teams are to a large extent autonomous, some functions are centralized, which may cause bottleneck problems. As an example, there is only one operator actively monitoring the ticket-management system, why it can take a long time before certain tickets are addressed. One interviewee exemplified how an individual had asked a followup question about a ticket he had reported some months earlier. The question was listed in place 80 in the ticket-management system. The ticket was assigned to a developer who then bounced it back to the operator with the question \textit{``Do we have to answer this?''}, further highlighting the closed mindset present in some of the teams. 


\vspace{3mm}
\noindent\textit{CH6: Fear of Presenting On-going Work.}
A reason for the closed mindset mentioned by multiple interviewees is a fear from developers of sharing what they are working on. An embarrassment of code quality, as well as language and coding conventions, is one aspect. Another aspect is that of being analyzed and questioned as SPES is under recurring attention from the press and politicians. Some developers view SPES's reputation as having a negative impact on their willingness to open up. One interviewee believes that if the unit were to have a better track record in terms of product and service delivery it would provide developers with better confidence. 

Another interviewee, on the other hand, considered the affiliation with SPES as something positive as it is a government agency people know about. The same interviewee also saw the fear among some developers grounded in uncertainty about how OSS and the collaborative methodology works.

Some developers wish for the process of opening up towards the ecosystem to happen at a slow pace to see what the needs are from the ecosystem's members. One interviewee questioned the need for having the whole backlog open and whether it would be enough to enable external reporting of issues.

\vspace{3mm}
\noindent\textit{CH7: Internal Communication and Collaboration.}
There is a wish for improved internal communication of the vision of the platform and ecosystem, as well as the collaborative development approach that the unit intends to adopt. Interviewees highlight a need for stronger enforcement from management level and that the team autonomy slows down the adoption process. Some teams more described as more ``open'' than others depending on the developers present and their previous experience of OSS.

\begin{figure*}[t!]
\centering
\includegraphics[width=1\textwidth]{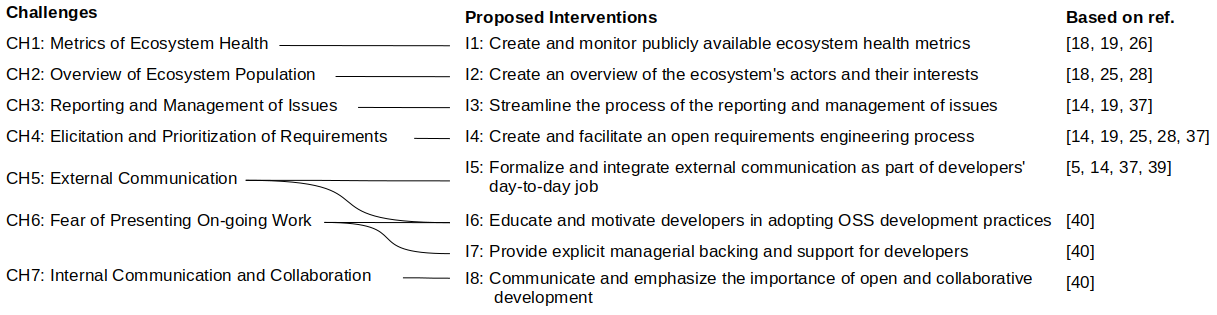}
\caption{Seven challenges (labeled CH\#) identified from the collected data and eight interventions (labeled I\#) proposed to address the different challenges based on the related work presented in section~\ref{sec:rw}.}
\label{fig:interventions}
\end{figure*}

Interviewees also request increased communication between the teams. Two teams have, for example, been implementing similar solutions but due to ``closed traditions'' as referred to by one interviewee, one team has remained independent in their work. Some teams are also reported using closed Slack channels for communication. One interviewee requests the possibility to present work and progress to others inside the unit to get feedback and learn from each other.

\section{Discussion and Proposed Interventions}
\label{discussion}
The identified challenges highlight that transforming an organization into a platform provider with an open collaborative development approach is not just about adopting certain practices. Cultural, as well as organizational and process-related aspects rooted within an organization, need to be addressed which can take time. Below we propose eight interventions (labeled I\#) for addressing the different challenges (see Fig~\ref{fig:interventions}) based on the related work presented in section~\ref{sec:rw}.









\vspace{3mm}
\noindent\textit{I1: Create and monitor publicly available ecosystem health metrics}.
A process-related aspect highlighted in CH1 is the lack of ecosystem health metrics. These metrics do not have to be explicitly listed but can, for example, be gained from open issue-trackers and roadmaps that can provide knowledge of the direction of the platform as well as the amount and type of issues~\cite{jansen2014measuring}. Such knowledge may enable actors to make decisions on whether to invest in the platform or how they should try to influence or contribute to it~\cite{linaaker2019community}. The unit is however recommended to take one step further and create explicit ecosystem health metrics and make these publicly available. This would be of value, not just for the ecosystem, but also for the unit itself, as it would allow them to monitor how the ecosystem evolves and provide input to how they should adapt their practices and the governance of the ecosystem~\cite{jansen2020focus}.

\vspace{3mm}
\noindent\textit{I2: Create an overview of the ecosystem's actors and their interests}.
As pointed out in CH2, one aspect of ecosystem health is knowing who the actors are, what niches they are focused on, and how these factors evolve~\cite{jansen2014measuring}. Knowledge about stakeholders and their agendas is also beneficial from a requirements engineering perspective as it may help in the elicitation and prioritization of requirements~\cite{linaaker2019method}. It can also help in facilitating and enabling an open and collaborative culture within the ecosystem as the actors can see who else is involved, which may also provide a picture of the platform's quality~\cite{manikas2013reviewing}.

\vspace{3mm}
\noindent\textit{I3: Streamline the process of the reporting and management of issues}.
A process-related aspect highlighted by in CH3 is the confusion of having multiple channels where bugs and feature requests could be reported. CH5 is an example of the frustration the reporting actors may feel. To address these challenges, the unit is recommended to adopt an open issue-tracker, which may be seen as a core practice in OSS development~\cite{scacchi2002understanding, ernst2012case}, while it may be considered a bit more mature practice for a general platform provider~\cite{jansen2020focus}. Issues or parts of the backlog, that are not considered relevant for the ecosystem can be kept on an internal backlog. An open by default approach should, however, be adopted even if ``double book-keeping'' will occur. Developers should actively use and monitor the issue-tracker, and also refer and instruct the ecosystem to use the issue-tracker instead of the many channels currently available.

\vspace{3mm}
\noindent\textit{I4: Create and facilitate an open requirements engineering process}.
Another process-related aspect identified in CH4 is that requirements have mostly been elicited internally within SPES. An open issue-tracker that is actively used would be one way to open up the requirements engineering of the platform and allow for the ecosystem to report but also help in the analysis and prioritization of the requirements~\cite{jansen2020focus, scacchi2002understanding}. Bug-reporting and quality assurance would also benefit from the collaborative approach as the ecosystem can, besides reporting, also partake in the triage, replication, and investigation of bugs~\cite{jansen2020focus}. 

Opening up the roadmaps would be another way to open up the requirements engineering process of the platform~\cite{jansen2020focus}. This would allow for the ecosystem to gain insight into the direction of the platform and provide feedback on how this aligns with their different agendas~\cite{linaaker2019method}. The JT-unit further needs to actively invite and engage the ecosystem into the requirements engineering process so that they start to assert requirements~\cite{manikas2013reviewing}, but also partake in their continuous evolution, similar to what is common in OSS ecosystems~\cite{scacchi2002understanding, ernst2012case}. 

\vspace{3mm}
\noindent\textit{I5: Formalize and integrate external communication as part of developers' day-to-day job}.
CH5 highlights a cultural aspect of external communication with the ecosystem, where similar patterns can be noticed as with the question of an open issue-tracker. Some developers consider it important and a natural part of their work, while others may view it as potentially cumbersome and interfering with their work. Education and encouragement may be needed along with clear processes and directives for how teams should communicate with the ecosystem. Providing quick responses, being welcoming and inclusive as well as pointing out good resources for newcomers to the ecosystem are some of the common best practices~\cite{steinmacher2019overcoming, barcomb2020managing}. Furthermore, the unit should consider their choices of both synchronous and asynchronous communication channels to ensure that the data is persistent, openly available and searchable, in order to support the decentralized aspects of open and collaborative development~\cite{scacchi2002understanding, ernst2012case}.

\vspace{3mm}
\noindent\textit{I6: Educate and motivate developers in adopting OSS development practices}.
A cultural aspect, highlighted specifically by in CH5-6, is the mindset among the developers which shifts between open and closed, depending on the team and their experience of OSS. Experienced developers, for example, see no problem with publishing code, being open, and collaborating with the ecosystem. Others, of which some may be less experienced of OSS, are more hesitant and request that the process of ``opening up'' is taken slowly. Education, motivation, and guidance on OSS development practices may, therefore, be warranted, for example, from internal OSS champions and management~\cite{stol2014key}.

\vspace{3mm}
\noindent\textit{I7: Provide explicit managerial backing and support for developers}.
Another cultural aspect highlighted by CH6 was that having an open issue-tracker was considered with skepticism by some as they feared being questioned by the press, due to the reputation of SPES and the attention it is given by media and politicians. Again, support from management may be a critical component~\cite{stol2014key}, and a clear message that the responsibility and accountability fall on the organization rather than on the individual.

\vspace{3mm}
\noindent\textit{I8: Communicate and emphasize the importance of open and collaborative development}.
CH7 puts a focus on both organizational and process-related aspects in terms of the internal communication and collaboration within the JT-unit. The unit should in this regard consider how to break the silos implied by the autonomous team-structure within the organization. Setting a clear policy may be one way to improve the internal communication of the vision of the ecosystem and about the open collaborative development approach~\cite{stol2014key}. Adoption of inner source practices such as a bazaar-style development and quality-assurance approach with standardized tooling and infrastructure between the teams may be a further step to help bridge the transition to an open collaborative development model~\cite{stol2014key}.


\section{Conclusions}
This case study explores how a public sector platform provider, aiming to create and grow an ecosystem around an underpinning platform of OGD and OSS, collaborates with the ecosystem on the platform development. Specifically, the study focus on what challenges they experience in their transformation towards an open and collaborative development model.

\emph{Challenges} highlight how developers fear being publicly questioned for their work, as their employer is a government agency, undergoing constant scrutiny. The requirements engineering is mainly a closed process, with no open issue trackers or roadmaps, nor actively facilitated towards the ecosystem, thereby preventing third parties from collaborating or influencing the platform development. There is no overview of the ecosystem's stakeholder population, why important relationships may risk not being maintained, sources of requirements being missed, and misinterpreted view of the ecosystem's health being created. 

Ecosystem health metrics are missing which makes it difficult to monitor ecosystem growth and effects from adopted practices and governance design. Also, it makes it complicated for existing and potential members of the ecosystem to decide whether to use (or keep using) the data and software provided by the platform. Communication with the ecosystem is managed through multiple channels causing internal and external confusion and is complicated further by a reluctance among some developers to communicate with third party. The internal autonomous team-structure is further reported to create silos making communication of the ecosystem's vision, enforcement of it, as well as general internal collaboration difficult.

We, therefore, conclude that public-sector platform providers must address both cultural, organizational, and process-related challenges to create and grow a healthy platform ecosystem. By also considering related work, we discuss and propose a number of potential \emph{interventions} that may be leveraged to address the different challenges. The validity and effect of these recommendations are topics for future work. Furthermore, considering the identified challenges, as this is an exploratory case study using qualitative research methods, we do not claim any statistical generalization. Readers should consider the context and unit of analysis of the reported case and adopt an analytical approach when attempting to transfer understanding to other cases~\cite{runeson2012casestudy}.

\section*{Acknowledgments}
We thank the interviewees for dedicating time and taking the courage to critically reflect on their current way of working. We also want to thank the three anonymous reviewers for providing constructive feedback that has helped to improve the paper. The research was funded by the Swedish Public Employment Service and their JobTech Dev program.

\bibliographystyle{ACM-Reference-Format}
\bibliography{references}

\appendix

\section{Questionnaire}\label{app:questionnnaire}
This section presents the questionnaire used in the semi-structured interviews. For each question, or area, the interviewee was asked to describe any related challenges or friction perceived by the interviewee, both internally and towards to ecosystem.
\begin{itemize}
    \item Describe your current process of testing your APIs and software released as OSS? Do you in any way involve the ecosystem in this process?
    \item Do you in any way support the ecosystem in maintaining and improving the quality of their implementations and integration with the platform?
    \item Describe how you initiate, grow, maintain, and manage your relationships with the ecosystem (both formally and informally)? 
    \item Describe in what ways (if any) you monitor the progress and evolution of the ecosystem and usage of the platform?
    \item Describe how you communicate and interact with the ecosystem? Differentiate between digital (synchronous and asynchronous communication) and physical interaction (e.g., meet-ups, hackathons, and conferences).
    \item Describe the development process of the software you are involved with? What possibilities does the ecosystem have to become involved in this process?
    \item Describe the infrastructure used to support the development process? In what ways (if any) does it support an open collaboration with the ecosystem? 
    \item Describe how source code and project artifact (e.g., documentation, roadmaps, backlog) (if any) are shared with the ecosystem?
    \item Describe in what ways (if any) you support the ecosystem in getting started with their development and use of the platform?
    \item Describe the process for reporting and managing tickets, including bugs, improvements, and features? In what ways (if any) can the ecosystem, be involved in this process?
    \item Describe your requirements engineering process for the platform, e.g., in terms of elicitation, prioritization, analysis, and specification? In what ways (if any) can the ecosystem, be involved in this process?
    \item Describe your roadmapping process? In what ways (if any) can the ecosystem, be involved in this process?
\end{itemize}

\end{document}